\documentstyle[12pt,epsfig]{article}

\begin{document}

\parindent 1.3cm
\thispagestyle{empty} 
\vspace*{-3cm}
\noindent

\def\aa{{\cal A}}
\def\lb{\overline\Lambda}
\def\kp{k_\perp}
\def\vp{v\!\cdot\! p}
\def\xt{\tilde x}
\def\BR{\rm BR}
\def\rb{\not \!\! R}

\def\be{\begin{equation}}
\def\ee{\end{equation}}
\def\ba{\begin{eqnarray}}
\def\ea{\end{eqnarray}}

\begin{obeylines}
\begin{flushright}
hep-ph/9803271
UG-FT-85/97
IC/98/13
March 1998
\end{flushright}
\end{obeylines}
\vspace{2cm}

\begin{center}
\begin{bf}
\centerline {MINIMAL SUPERSYMMETRIC SCENARIOS FOR}
\centerline {SPONTANEOUS ${\bf CP}$ VIOLATION}
\end{bf}
\vspace{1.cm}

M. Masip 

\vspace{0.1cm}
{\it Departamento de F\'\i sica Te\'orica y del Cosmos\\
Universidad de Granada\\
18071 Granada, Spain\\}

\vspace{1.cm}

A. Ra\v{s}in 

\vspace{0.1cm}
{\it High Energy Section\\ 
ICTP\\
34014 Trieste, Italy\\}
\vspace{2.2cm}

{\bf ABSTRACT}

\parbox[t]{12cm}{
We study the possibility of spontaneous $CP$ violation (SCPV)
at the tree level in models with an extended Higgs sector. We show that
the minimum equations for the complex phases of the vacuum 
expectation values (VEVs) have always a geometrical 
interpretation in terms of triangles.  To illustrate our
method  we analyze the minimal
supersymmetric (SUSY) model with R-parity violating couplings and 
sneutrino VEVs, where there is no SCPV. 
Then we study SUSY models with extra 
Higgs doublets and/or gauge singlets, and 
find that the simplest scenario with SCPV must include
at least two singlet fields. 
}
\end{center}
\newpage

\section{Introduction} 

Although the observed $CP$ violation can easily be accommodated 
in the standard model, allowing for arbitrary 
$CP$ violating terms leads to phenomenological difficulties 
in models with an enlarged Higgs sector.
For example, in general two-Higgs doublet 
models with acceptable flavor changing interactions the prediction for  
$\epsilon_K$ would still be too large \cite{hall93}. 
Another example is the minimal SUSY
extension of the standard model (MSSM), where arbitrary 
complex phases in gaugino masses and scalar trilinears would produce 
too large electric dipole moments \cite{deru90}.

In order to bring the predicted $CP$ violation in such extensions 
to acceptable levels one can consider imposing additional symmetries, 
like a flavor symmetry or a discrete symmetry in 
the Higgs sector. One obvious possibility is to impose $CP$ 
invariance itself. If $CP$ violation 
is spontaneous \cite{leet73}, in the sense that it appears via VEVs 
of scalar fields, then some phases could be naturally suppressed 
by the ratio of two mass scales present in the model, 
like the top and bottom quark masses \cite{babu94,masi95a} 
or the electroweak and the SUSY scales. 
This hierarchy could be used, for example, to accommodate small 
complex phases in gaugino masses and scalar trilinears together with 
the large CKM phase required to explain the kaon system. 

Another interesting motivation for SCPV arises 
from the strong $CP$ problem. As far as $CP$ is a good symmetry the
QCD phase $\theta$ is zero.  The spontaneous breaking
of $CP$ could respect this initial value of $\theta$
while creating the observed weak $CP$ violation \cite{geor78,barr79}.
Such scenarios find a natural framework in left-right models,
where it was recently shown that a minimal model of this type must be
supersymmetric and with a low scale of $SU(2)_R$ symmetry breaking  
\cite{moha97}.

In consequence, SCPV appears as a well motivated possibility in 
SUSY models. A priori, these models contain
enough ingredients and arbitrarity to introduce SCPV: 
a neutral scalar sector with at least 
two Higgs fields plus three sneutrinos,
and a large number of arbitrary SUSY-breaking terms.
However, it is well known that 
the MSSM has only real minima.
In fact, complex VEVs are possible once radiative
corrections are included \cite{maek92}, but then the model contains a
too light Higgs boson which is experimentally excluded \cite{poma92a}.
Therefore, it is necessary to introduce other fields
and couplings. Two cases have been considered before: an extra
gauge singlet field (2D1S model) 
and an extra pair of Higgs doublets (4D model). 
The presence of a singlet is appealing because it defines
a scenario where the SUSY mass term of
the Higgs is substituted by a trilinear 
term in the superpotential
combined with a singlet VEV, avoiding the 
$\mu$ problem and relaxing the upper bound on the mass of the
lightest neutral Higgs \cite{masi98}. 
The singlet model (known as the next--to--MSSM) has been 
analyzed in this context by Romao \cite{roma86}, 
with the result that no $CP$ violating minima
exist at tree level\footnote{We do not consider here singlet extensions
where dimensionful couplings are allowed. In such case it was shown
that spontaneous $CP$ violation was possible \cite{poma92}.}.
The inclusion of an extra pair of Higgs doublets is also an obvious
generalization of the MSSM. But the result there is negative as 
well, the minimum in the 4D model is always real \cite{masi95}.
In Section 4 we review the arguments used to prove these results. 

Contrary to non-SUSY models, where the
Higgs quartic interactions are arbitrary, 
SCPV in SUSY scenarios seems to require an increasing number 
of species and fields. This fact quickly will make the analysis of
minimum equations cumbersome. In Section 2 we describe a 
method to find complex minima that interprets the equations
for the phases in terms of a geometrical object 
combining triangles. 
The method is general in the sense that it
can be applied to any potential (SUSY or not). We illustrate 
it analyzing a SUSY model with minimal matter content but
R-parity breaking couplings and sneutrino VEVs.
Our objective in this paper is to find the minimal SUSY scenario 
giving SCPV in the Higgs sector at the tree level. 
In Section 3, we extend the Higgs sector of 
the MSSM with singlets and/or extra doublets
(the addition of $SU(2)_L$ triplets is 
phenomenologically disfavored), and we 
show that the minimal scenario consists of at least two
singlets, regardless of the number of Higgs doublets.
In Section 4 we give our final remarks and conclusions.

\section{The geometrical method}

In this section we describe a geometrical method of 
analyzing the minima equations for the phases that was recently 
used in the context of a SUSY model with four Higgs doublets. 
The method is actually a generalization of the procedure used
in the simplest case, where the object representing the  
equations is just a triangle \cite{roma86,bran80}. Let us start
studying the non-SUSY model proposed in \cite{wein76} 
to review this simplest case, and then we will 
describe in some detail a more general scenario.

Consider the extension of the standard model with three Higgs
doublets \cite{wein76} where one doublet couples to the up quarks, 
another one to the down quarks, and the third one does not couple 
to quarks at all. The scalar potential is given by
\ba
V_{3D} &=& m^2_i H^\dagger_i H_i + \lambda_i (H^\dagger_i 
H_i)^2 + \lambda'_{ij} (H^\dagger_i H_j) (H^\dagger_j H_i) 
\nonumber\\
&&+ [ \lambda_{12} (H^\dagger_1 H_2) (H^\dagger_1 H_2) +
\lambda_{13} (H^\dagger_1 H_3) (H^\dagger_1 H_3) 
\nonumber\\
&&+ \lambda_{23} (H^\dagger_2 H_3) (H^\dagger_2 H_3)  + {\rm h.c.} ]\;,
\ea 
with all the parameters real. The neutral components of the doublets
will have VEVs
\be
\langle H_i \rangle =
{1 \over \sqrt{2}} v_i e^{i\delta_i} 
\;\;\;\; (i=1,2,3)\;,
\ee
where an hypercharge transformation is used  to set 
$\delta_1 = 0$ (the minimum will be degenerate due to the $U(1)_Y$
symmetry of the potential). The value of $v_i$ and $\delta_i$ will
be given by the solution of 
the minimum equations. For the complex phases these are
(${\partial V \over \partial \delta_i} = 0 $) \cite{desh77}
\ba
\lambda_{12} v_1^2 v_2^2 \sin\delta_2 +
\lambda_{23} v_2^2 v_3^2 \sin(\delta_2 + \delta_3) &=& 0
\nonumber\\ 
\lambda_{23} v_2^2 v_3^2 \sin(\delta_2 + \delta_3) +
\lambda_{13} v_1^2 v_3^2 \sin\delta_3 &=& 0
\ea
To solve these equations we draw a triangle \cite{bran80} with
sides $a_1^{-1}$, $a_2^{-1}$, $a_3^{-1}$ and opposite angles
$\pi-\delta_2$, $\pi-\delta_3$, $\delta_2+\delta_3-\pi$, 
respectively (this is the lower right triangle in Fig.~1).
If we define
\ba
a_1 & = & \lambda_{12} v^2_1 v^2_2 
\nonumber\\
a_2 & = & \lambda_{13} v^2_1 v^2_3 
\nonumber\\
a_3 & = & \lambda_{23} v^2_2 v^2_3 
\ea
then the sine law applied to the triangle implies Eq.~(3), i.e.,
the triangle is the solution to Eq.~(3).
Using the triangle it is now possible to give the values of the cosines 
of the phases (which appear in the minimum equations for the
moduli) in terms of the sides:
\ba
\cos\delta_2 & = & 
{ {(\lambda_{32}v^2_3v^2_2)(\lambda_{13}v^2_1v^2_3)} \over 2}
[({1 \over {\lambda_{12}v^2_1v^2_2}})^2
-({1 \over {\lambda_{32}v^2_3v^2_2}})^2-
({1 \over {\lambda^2_{13}v^2_1v^2_3}})^2]
\nonumber\\
\cos\delta_3 & = & 
{ {(\lambda_{32}v^2_3v^2_2)(\lambda_{12}v^2_1v^2_2)} \over 2}
[({1 \over {\lambda_{13}v^2_1v^2_3}})^2
-({1 \over {\lambda_{32}v^2_3v^2_2}})^2-
({1 \over {\lambda^2_{12}v^2_1v^2_2}})^2]
\nonumber\\
\cos(\delta_2+\delta_3) & = & 
{ {(\lambda_{12}v^2_1v^2_2)(\lambda_{13}v^2_1v^2_3)} \over 2}
[({1 \over {\lambda_{32}v^2_3v^2_2}})^2
-({1 \over {\lambda_{12}v^2_1v^2_2}})^2-
({1 \over {\lambda^2_{13}v^2_1v^2_3}})^2]\nonumber\\
\label{coswein}
\ea
Substituting these expressions in the three minimum equations
for the moduli we will obtain the equations in terms of the
three moduli $v_i$ only (with no phases), 
and these equations can be solved numerically. 
The value of the moduli will fix the sides of the
triangle and, in consequence, the value of the complex phases.
For the particular three doublet model under study, it is 
easy to find solutions $v_i$ for adequate values of the couplings,  
and then there is SCPV. Below we analyze 
a similar example where the phases can also be expressed
in terms of the moduli, but then the minimum equations for the
moduli are always incompatible.

This method of expressing the phases as functions of the moduli
can be generalized to models with more than two complex phases.
We will show now how to build a set of triangles, with their angles 
related to the phases in the VEVs, that solves the phase minimum 
equations. The method can be used if 
the equations involve only sines (and not cosines) of the phases,
which is the case for Higgs potentials with all the 
couplings real. 
In models with $CP$ odd fields the $CP$ invariant potential may
include purely imaginary couplings. 
However, since in SUSY models all the scalar fields are complex, 
the factors of $i$ in the couplings 
can always be rotated away by field redefinitions. 
In non-SUSY cases there may be exceptions with $CP$ odd real 
scalars\footnote{We thank Goran Senjanovi\'{c} for pointing out 
this posssibility.}. The (real) VEVs of such fields would 
break $CP$, which would be transmitted to the rest
of the Lagrangian through complex couplings either
through additional fermions \cite{dval96}  
or through explicit couplings to the Higgs doublets. The 
minimal model of the latter type 
is a two Higgs doublet model with flavor conserving 
couplings \cite{glas77} and with one $CP$ odd real singlet. In this 
case the minimum of the potential has terms $\sin(\delta_2)$ 
and $\cos(2\delta_2)$, being $\delta_2$ the relative phase of 
the two doublets. Here, however, we can also do a  
rotation of $\delta_2$ by $\pi/2$ to convert both terms into cosines,
and the minimum equation will only involve sines of the phase. 

To describe our method in some detail we will 
consider an extension of the MSSM with 
$R$ parity violating couplings ($\rb$ model). 
The model has a minimal matter content with one
pair of Higgs doublets. However, sneutrino VEVs 
are allowed. This fact will introduce, in an
effective way, three new doublets of $-1/2$ hypercharge
in the scalar Higgs sector.
The relevant part of the superpotential is  
\be
W = W_{MSSM} + \mu_i\; H_2 L_i 
\ee
We will restrict our analysis to the VEVs 
of the Higgs ($H_1,H_2$) and lepton ($L_i$) fields, 
assuming that we are in 
the charge conserving part of the parameter space. We will closely
follow the notation used in the multi-Higgs 
case in \cite{masi95}, with odd (even) indices indicating
$-1/2$ ($+1/2$) hypercharge. We redefine $(L_1,L_2,L_3)
\rightarrow (H_3,H_5,H_7)$ and write the VEVs for the neutral
components of these doublets as
\ba
\langle H_i\rangle & = & 
{1 \over \sqrt{2}} v_i e^{i\delta_i} 
\;\;\;\; (i=1,2,3,5,7)\;,
\ea
using a global hypercharge transformation to set $\delta_1 = 0$.
Then the VEV of the scalar potential is
\begin{eqnarray}
V_{\not \! R}& = & 
{1 \over 2} m^2_1 v^2_1 + 
{1 \over 2} m^2_2 v^2_2 + 
{1 \over 2} m^2_3 v^2_3 + 
{1 \over 2} m^2_5 v^2_5 + 
{1 \over 2} m^2_7 v^2_7  + 
m^2_{12} v_1 v_2 \cos\delta_2\nonumber\\
&&+ 
m^2_{13} v_1 v_3 \cos\delta_3+
m^2_{15} v_1 v_5 \cos\delta_5+
m^2_{17} v_1 v_7 \cos\delta_7+
m^2_{32} v_3 v_2 \cos(\delta_3+\delta_2)\nonumber\\
&&+
m^2_{52} v_5 v_2 \cos(\delta_5+\delta_2)+
m^2_{72} v_7 v_2 \cos(\delta_7+\delta_2)+
m^2_{35} v_3 v_5 \cos(\delta_3-\delta_5)\nonumber\\
&&+
m^2_{37} v_3 v_7 \cos(\delta_3-\delta_7)+
m^2_{57} v_5 v_7 \cos(\delta_5-\delta_7) + V_D\nonumber\\
\end{eqnarray}
where all the mass parameters are real and
$V_D$ is the D-term part of the potential 
\be
V_D = {1 \over 32} (g^2 + g'^2) [v^2_1 + v^2_3 + v^2_5 + v^2_7 - 
v^2_2]^2
\ee

We are seeking $CP$-violating minima, i.e., 
minima where some phases are different from $0$ or $\pi$.
First, we solve the minimum conditions for the 
phases ($- {{\partial V} \over {\partial \delta_i}} = 0$):
\ba
&  & 
m^2_{12} v_1 v_2 \sin\delta_2 +
m^2_{32} v_3 v_2 \sin(\delta_3+\delta_2) +
m^2_{52} v_5 v_2 \sin(\delta_5+\delta_2) +
m^2_{72} v_7 v_2 \sin(\delta_7+\delta_2)  
= 0 \nonumber\\
&  & 
m^2_{32} v_3 v_2 \sin(\delta_3+\delta_2) +
m^2_{13} v_1 v_3 \sin\delta_3 +
m^2_{35} v_3 v_5 \sin(\delta_3-\delta_5) +
m^2_{37} v_3 v_7 \sin(\delta_3-\delta_7)
= 0 \nonumber\\
&  & 
m^2_{52} v_5 v_2 \sin(\delta_5+\delta_2) -
m^2_{35} v_3 v_5 \sin(\delta_3-\delta_5) +
m^2_{15} v_1 v_5 \sin\delta_5 +
m^2_{57} v_5 v_7 \sin(\delta_5-\delta_7)
= 0 \nonumber\\
&  & 
m^2_{72} v_7 v_2 \sin(\delta_7+\delta_2) -
m^2_{37} v_3 v_7 \sin(\delta_3-\delta_7) -
m^2_{57} v_5 v_7 \sin(\delta_5-\delta_7) +
m^2_{17} v_1 v_7 \sin\delta_7 
= 0 \nonumber\\
\label{phasemin}
\ea
Now we express the solution
of the minimum equations for the phases in terms of a combination
of triangles. We observe that there are ten independent quantities
$m_{ij}v_iv_j$ in the equations and, in consequence, the space of 
solutions
will be 10-dimensional. The geometrical solution must be
given by a set of six triangles whose angles involve the different 
combinations of four phases in 
Eq.~(\ref{phasemin}). Such a combination of triangles
contains ten independent sides\footnote{
Note that the four angles and one side of each triangle 
also fixes the six triangles. In general, the number of
phases in the equations plus the number of triangles in
the multitriangle must be equal to the number of independent 
parameters in the equations.}.
A choice for the six triangles is shown in Fig.~1.
Adding the sine law applied to the triangles it is straightforward to
obtain the set of equations in (\ref{phasemin}), where
we identify 
\begin{eqnarray}
&&m_{12}^2 v_1 v_2=a_1+b_1+x_1\;,\;\;
m_{32}^2 v_3 v_2=a_3\;,\;\;
m_{52}^2 v_5 v_2=-b_3\;,\;\;\nonumber \\
&&m_{13}^2 v_1 v_3=a_2+c_1+z_1\;,\;\;
m_{72}^2 v_7 v_2=-x_3\;,\;\;
m_{35}^2 v_3 v_5=-c_3\;,\;\;\nonumber \\
&&m_{15}^2 v_1 v_5=b_2-c_2+y_1\;,\;\;
m_{37}^2 v_3 v_7=-z_3\;,\;\;
m_{57}^2 v_5 v_7=-y_3\;.\;\;\nonumber \\
&&m_{17}^2 v_1 v_7=x_2-z_2-y_2\;,\;\;
\label{s0}
\end{eqnarray}
Now we have to express $\cos \delta_i$ in terms of $m_{ij}v_iv_j$
and substitute them in the five minimum equations for the moduli. 
If there we find a solution fixing $v_i$, then we have SCPV; 
if the equations are incompatible then any minimum
that may exist will be $CP$ conserving. 
Although this procedure sounds simple, in general it is not easy to
obtain analytic solutions. 
In particular,  to express the sides of the triangles in 
terms of $m_{ij}v_iv_j$ involves solving a 
quartic equation. The multitriangle is useful to generate numerical 
solutions, as we do in Section 3 to study other models, but
it is not enough to solve analytically complicated cases. 
In the $\rb$ model under study here, however, 
we can find a way to simplify
the equations that allows for analytical solutions.
We can rotate the original fields 
($H_1$,$H_3$,$H_5$,$H_7$) to a new basis so that
\be 
m^2_{13}=m^2_{15}=m^2_{17}=m^2_{35}=m^2_{37}=m^2_{57}=0\;.
\ee
This rotation redefines the diagonal terms $m^2_i$ but does not 
introduce any complex phase in the Lagrangian. From 
Eq.~(\ref{phasemin}) this set of zero masses forces
\be
\sin\delta_2 
= \sin(\delta_2+\delta_3)
= \sin(\delta_2+\delta_5)
= \sin(\delta_2+\delta_7) = 0
\ee
which, if the five VEVs $v_i$ are non-zero, 
forces all the phases to be trivial. Then there 
is no $CP$ violation in the general case with all $v_i\not= 0$.

If the modulus
$v_7 =0$ and all the other VEVs nonzero, then the phase $\delta_7$ 
is irrelevant 
(from (\ref{s0}) we see that $x_3=y_3=z_3=0$ and the corresponding 
triangles become infinite). Now the three-phase case
($\delta_2,\delta_3,\delta_5$) 
that results has three triangles left and is 
completely analogous to the 4D model 
discussed in \cite{masi95}, with  no SCPV.

The case with two VEVs zero, $v_5=v_7=0$, appeared
in \cite{nowa94}, where it is 
claimed that there are  $CP$-violating minima.
Since we disagree with this result,
let us consider it in a bit more detail. Now the minimum
equations for the phases reduce to
\ba
&  & 
m^2_{12} v_1 v_2 \sin\delta_2 +
m^2_{32} v_3 v_2 \sin(\delta_3+\delta_2) 
= 0 \nonumber\\
&  & 
m^2_{32} v_3 v_2 \sin(\delta_3+\delta_2) +
m^2_{13} v_1 v_3 \sin\delta_3 
= 0
\ea

Again, the lower right triangle in Fig.~1 solves these 
equations if 
$a_1 = m_{12} v_1 v_2$, 
$a_2 = m_{13} v_1 v_3$ and 
$a_3 = m_{23} v_2 v_3$. The cosines of the angles
can be obtained from the expressions in (\ref{coswein}) just 
exchanging 
$ \lambda_{ij}v^2_iv^2_i \rightarrow m^2_{ij}v_iv_j$.

Notice that this does {\it not} mean yet that there is
SCPV, we still have to solve the three minimum 
equations for
the moduli ($v_i {{\partial V} \over {\partial v_i}} =0$):
\ba
m^2_1 v^2_1 +
m^2_{12} v_1 v_2 \cos\delta_2 +
m^2_{13} v_1 v_3 \cos\delta_3 + v^2_1 g({\bf v})
&=& 0 \nonumber\\
m^2_2 v^2_2 +
m^2_{12} v_1 v_2 \cos\delta_2 +
m^2_{23} v_2 v_3 \cos(\delta_2+\delta_3) - v^2_2 g({\bf v})
&=& 0 \nonumber\\
m^2_3 v^2_3 +
m^2_{13} v_1 v_3 \cos\delta_3 +
m^2_{23} v_2 v_3 \cos(\delta_2+\delta_3) + v^2_3 g({\bf v})\;,
&=& 0 \nonumber\\
\ea
where $g({\bf v}) \equiv {1 \over 8} (g^2+g'^2) (v_1^2+v_3^2-v^2_2)$.
We substitute the expressions for $\cos\delta_i$ in the 
equations above and obtain 
\ba
v^2_1 [ m^2_1 - { {m^2_{12} m^2_{13}} \over m^2_{32}} + g({\bf v}) ]
&=& 0 \nonumber\\
v^2_2 [ m^2_2 - { {m^2_{32} m^2_{13}} \over m^2_{12}} - g({\bf v}) ]
&=& 0 \nonumber\\
v^2_3 [ m^2_3 - { {m^2_{32} m^2_{12}} \over m^2_{13}} + g({\bf v}) ]
&=& 0 \nonumber\\
\ea
There is only one combination of VEVs, $g({\bf v})$, to solve the
system of three equations. Then, unless the masses are fine tuned 
in such a way that two of the moduli equations are trivial, the three 
equations can not be satisfied simultaneously.
This implies that the minimum equations for the phases
must have the trivial solution
(phases equal zero or $\pi$). 
We conclude that there is no tree-level SCPV in the
extension of the MSSM 
with R-parity violating couplings.

In the two simple non-SUSY and SUSY cases above 
the phase minimum equations have been solved in terms of
triangles. It is now possible to see how to extend the procedure to 
more complicated cases. If we add a Higgs field to any of
the previous models, this field will introduce one
more phase and will increase the number of independent  
parameters in the equations. To find the solution now 
we need to draw new triangles, and the number of independent
distances in these triangles must be exactly equal to the number
of new parameters (couplings) in the equations. This will be 
always possible because we can draw an arbitrary
number of new triangles with the existing phases, and each new triangle
will introduce one more independent distance: the overall scale 
of the triangle, which is not fixed by the phases. We 
can always add triangles until matching the number of new couplings
in the minimum equations.
The procedure will become more obvious in the next Section, where
we consider SUSY models with additional Higgs fields.

\section{Search for the minimal SUSY model} 

The method described in the previous section has been already
applied to minimal SUSY scenarios, namely, to the 2D1S model
and to the 4D model. In both cases the results 
are negative. In the 2D1S model the 
minimum equations for the phases are also solved by a single triangle. 
When this triangle is imposed on the equations
for the moduli it was found 
\cite{roma86} that the solution is always a
saddle point: the Hessian matrix has
a negative eigenvalue. The scalar 
mass matrix is given by the second derivatives
of the potential, and this negative eigenvalue is equivalent to
the negative mass squared distinctive of {\it false}
vacua. One may rely on radiative corrections to turn 
the negative mass into positive \cite{babu94a}, but even then the 
model gives a field which is too light to have escaped detection. 
In consequence, there is no SCPV in the singlet model.

In the 4D model the equations for the phases are solved
by three triangles. When this solution is imposed
on the four minimum equations for the moduli we found 
\cite{masi95} that they only depend on 
two combinations of VEVs and, in consequence, there is 
no solution. A solution can be obtained if the
mass parameters of the scalar potential are fine tuned 
and two of the four equations become trivial (just like
in the model studied in Section 2). 
But even this fine tuned solution is not phenomenologically
acceptable, because it is degenerate
and predicts two massless fields. Radiative corrections
would relax the required amount of fine tuning and would
give mass to all the fields, but still the model 
has two particles which seem to be too light \cite{masi95a}. 
Therefore, there is no SCPV in the 4D model neither. 
The $\rb$ model discussed in section II is in some way a particular 
case of the 4D model, with an analogous negative result.

In this section we explore further extensions of the MSSM. 
We shall consider the models with three pairs of Higgs doublets
(6D model), with two pairs of doublets plus one singlet 
(4D1S model) and with two singlets (2S model). 
When adding singlets, we
will not include in the superpotential any couplings with
dimensions of mass:  $\mu$ terms for the doublets 
or linear and bilinear terms
for the singlets.  When such a term does not appear in the superpotential,
we will not include the corresponding soft SUSY breaking term in
the scalar potential neither (we assume that the SUSY breaking
mechanism respects the discrete symmetries of the superpotential).

The scalar potential for the neutral fields is in each case
\begin{eqnarray}
V_{6D} &=&  \sum_{i=1}^{6} m_i^2\; H^\dagger_i H_i 
+ (\; m_{13}^2\; H^\dagger_1H_3 + m_{15}^2\; H^\dagger_1H_5 + 
m_{35}^2\; H^\dagger_3H_5 + {\rm h.c.}\;) \nonumber \\
&&+ (\; m_{24}^2\; H^\dagger_2H_4 + m_{26}^2\; H^\dagger_2H_6 + 
m_{46}^2\; H^\dagger_4H_6 + {\rm h.c.}\;) \nonumber \\
&&+ (\; m_{12}^2\; H_1H_2 + m_{14}^2\; H_1H_4 + 
m_{16}^2\; H_1H_6 + m_{32}^2\; H_3H_2 + m_{34}^2\; H_3H_4 
\nonumber \\
&&+ m_{36}^2\; H_3H_6 + m_{52}^2\; H_5H_2 + 
m_{54}^2\; H_5H_4 + m_{56}^2\; H_5H_6 + {\rm h.c.}\;) 
\nonumber \\
&&+ {1\over 8}(g^2+g'^2)[H_1^2+H_3^2+H_5^2-H_2^2-H_4^2-H_6^2]^2\;,
\nonumber \\
\label{v6d}
\end{eqnarray}
\begin{eqnarray}
V_{4D1S} &=&  \sum_{i=1}^{4} m_i^2\; H^\dagger_i H_i 
+ m_5^2\; S^\dagger S 
+ (\; m_{13}^2\; H^\dagger_1H_3 + 
m_{24}^2\; H^\dagger_2H_4 + {\rm h.c.}\;) \nonumber \\
&&+ (\; \beta_{12}\; SH_1H_2 + \beta_{14}\; SH_1H_4 + 
\beta_{32}\; SH_3H_2 + \beta_{34}\; SH_3H_4 
\nonumber \\
&&+ {\beta_{5}\over 2}\; S^3 + {\rm h.c.}\;) 
+ \mid \alpha_{12}\; SH_2 \mid^2 + 
\mid \alpha_{12}\; SH_1 \mid^2 + \mid \alpha_{34}\; SH_4 \mid^2 
\nonumber \\
&&+ \mid \alpha_{34}\; SH_3 \mid^2 + \mid \alpha_{12}\; H_1H_2 + 
\alpha_{34}\; H_3H_4 + 
\lambda\; S S \mid^2 \nonumber \\
&&+ {1\over 8}(g^2+g'^2)[H_1^2+H_3^2-H_2^2-H_4^2]^2\;,
\nonumber \\
\label{v4d1s}
\end{eqnarray}
and 
\begin{eqnarray}
V_{2S} &=& \sum_{i=1}^{4} m_i^2\; S^\dagger_i S_i 
+ {1\over 2}\; (\; \beta_{3}\; S^2_1S_2 + \beta_{4}\; S_2^2 S_1+ 
\beta_{1}\; S_1^3 + \beta_{2}\; S_2^3 + {\rm h.c.}\;) \nonumber \\
&&+ \mid \alpha_{3}\; S_1S_2 +
{\alpha_{4}\over 2}\; S_2^2 + \alpha_{1}\; S_1^2 \mid^2 
+ \mid {\alpha_{3}\over 2}\; S_1^2 + \alpha_{4}\; S_2S_1 +
\alpha_{2}\; S_2^2 \mid^2 \;.\nonumber \\
\label{v2s}
\end{eqnarray}
In the $4D1S$ model we have rotated the doublets in the superpotential
so that $\alpha_{14}=\alpha_{32}=0$. 
In the $2S$ model we shall consider for simplicity
only the singlet sector, since this will be enough to prove that
there is SCPV.

In the 6D model we search for a complex minimum of type
\begin{equation}
\langle H_i \rangle = {1\over \sqrt 2} v_i e^{i\delta_i}
\;\;\;(i=1,6)\;,
\label{vev1}
\end{equation}
with $\delta_1=0$. The minimum conditions for the
phases $\delta_i$ give five equations. 
In these equations there appear 15 combinations of
masses and moduli $m_{ij}^2 v_i v_j$. Following the procedure
described in Section 2 we find that the geometrical solution 
consists of the ten triangles in Fig.~(2)\footnote{In a 
generic SUSY model with 2n Higgs doublets one
can choose (n-1)(2n-1) triangles with all possible pairs of phases.
These will have n(2n-1) independent sides, equal to the
number of $m^2_{ij}$ that appear in the equations.}. 

A given value of $m_{ij}^2 v_i v_j$ fixes
the multitriangle solution: 
\begin{eqnarray}
&&m_{12}^2 v_1 v_2=a_1-b_1-x_1+y_1\;,\;\;
m_{13}^2 v_1 v_3=a_2-c_2-w_1+e_2\;,\;\;
m_{32}^2 v_3 v_2=a_3\;,\;\;\nonumber \\
&&m_{14}^2 v_1 v_4=x_2-c_1-d_1+z_1\;,\;\;
m_{15}^2 v_1 v_5=f_2-b_2-d_2+w_2\;,\;\;
m_{24}^2 v_2 v_4=x_3\;,\;\;\nonumber \\
&&m_{16}^2 v_1 v_6=f_1-y_2-z_2+e_1\;,\;\;
m_{52}^2 v_5 v_2=b_3\;,\;\;
m_{34}^2 v_3 v_4=c_3\;,\;\;
m_{26}^2 v_2 v_6=y_3\;,\;\;\nonumber \\
&&m_{36}^2 v_3 v_6=e_3\;,\;\;
m_{35}^2 v_3 v_5=w_3\;,\;\;
m_{46}^2 v_4 v_6=z_3\;,\;\;
m_{54}^2 v_5 v_4=d_3\;,\;\;
m_{56}^2 v_5 v_6=f_3\;,\;\;\nonumber \\
\label{s1}
\end{eqnarray}
We can now choose a particular 10-triangle and find
the mass parameters which correspond to that solution. Once we
have these parameters 
we find the second derivatives of the potential
to check that the solution is indeed 
a minimum. It turns out that the minimum is always degenerate:
any 10-triangle solution occurs for a set of mass 
parameters giving two massless eigenstates (in addition to the 
goldstone boson of the hypercharge). The
situation here is 
completely analogous to the 4D model. The minimum 
equations have in general no solution. A solution is obtained
only for a fine tuned choice of mass parameters that
renders two of the six moduli equations trivial. 
Then there appear the distinctive 
two massless eigenstates. We have generated numerically 
random 10-triangle solutions and have obtained always the
same type of nonacceptable minimum. In consequence, we conclude
that there is no $SCPV$ in 4D and 6D models: complex
minima will require the presence of singlet fields. 

Next we study the 4D1S case. The VEVs can be written
\begin{eqnarray}
\langle H_i \rangle &=& 
{1\over \sqrt 2} v_i e^{i\delta_i}\;\;\;(i=1,4)\;,\nonumber \\
\langle S \rangle &=& v_5 e^{i\delta_5}\;,\nonumber \\
\label{vev2}
\end{eqnarray}
with $\delta_1=0$. The minimum conditions for the
phases $\delta_i$ define three equations. It is convenient to
rename $(\delta_2+\delta_5)\rightarrow \delta_2$,
$(\delta_4+\delta_5)\rightarrow \delta_4$, and
$3\delta_5\rightarrow \delta_5$, and
$\delta_3$ remains unchanged.
In Fig.~3 we plot the combination of triangles that solves 
the 5 equations. The sides in these triangles are related to
the parameters of the scalar potential:
\begin{eqnarray}
&&\beta_{12} v_1 v_2 v_5=a_1+b_1+x_2\;,\;\;
\beta_{32} v_3 v_2 v_5=a_3\;,\;\;
m_{24}^2 v_2 v_4=y_3-b_3\;,\nonumber \\
&&\beta_{34} v_3 v_4 v_5=z_1-c_3\;,\;\;
\beta_{14} v_1 v_4 v_5=c_2-b_2\;,\;\;
m_{13}^2 v_1 v_3=a_2-y_1+c_1\;,\nonumber \\
&&\alpha_{12}\lambda v_1 v_2 v_5^2=
-x_3\;,\;\;
\alpha_{34}\lambda v_3 v_4 v_5^2=
z_3\;,
\nonumber \\ 
&&{1\over 2}\alpha_{12}\alpha_{34} v_1 v_2 v_3 v_4=
-y_2\;,\;\;
\beta_{5} v_5^3=-x_1-z_2\;.
\label{s2}
\end{eqnarray}
The six triangles in Fig.~3 depend on the ten independent 
distances above. We proceed like in the 6D model, choosing
numerically 
a particular multitriangle solution and adjusting the
parameters in the potential in order to have that minimum.
Then we check if the point is really a minimum and we find its
properties.
We obtain that, although the point given by this method has
zero first derivatives, it is never a minimum. Like in the
singlet model analyzed by Romao \cite{roma86}, 
here the Hessian has the 
negative eigenvalues that characterize a saddle point. For all
the (random) cases that we have produced we find negative
eigenvalues, never a minimum. We conclude that the extension of the
MSSM with one singlet plus an extra pair of doublets does not
offer the possibility of SCPV neither.

Let us finally consider the 2S model in Eq.~(\ref{v2s}). We search for
complex minima of type
\begin{equation}
\langle S_i \rangle = v_i e^{i\delta_i}\;\;\;(i=1,2)\;.
\label{vev3}
\end{equation}
The two minimum equations for the phases are solved by the 
set of four triangles in Fig.~4, where the six independent
distances in the triangles are related to the six independent
parameters in the scalar potential:
\begin{eqnarray}
&&\beta_{3} v_1^2 v_2 =-c_1\;,\;\;
\beta_{4} v_1 v_2^2=-a_2\;,\;\;
3\beta_{1} v_1^3=2c_2+a_3+c_2{x_3\over c_1}-a_3{b_1\over a_2}
\;,\nonumber \\
&&3\beta_{2} v_2^3=b_2-x_1+b_2{c_3\over b_3}+x_1{2a_1\over x_2}
\;,\nonumber \\
&&v_1v_2(\alpha_3\alpha_4v_2^2+2\alpha_1\alpha_3 v_1^2+
\alpha_3\alpha_4v_1^2+2\alpha_2\alpha_4 v_2^2)=b_3
\;,\nonumber \\
&&2v_1^2v_2^2(\alpha_1\alpha_4+\alpha_2\alpha_3)=-d_2
\;.\nonumber \\
\label{s3}
\end{eqnarray}
 From the triangles we can construct the solution with 
$v_1=1$, $v_2=1.5$, $\theta_1=\pi/6$ and $\theta_2=\pi/12$.
This minimum corresponds to $\beta_1=0.61$, 
$\beta_2=0.05$, $\beta_3=-0.33$, $\beta_4=-0.13$, 
$\alpha_1=2.8$, 
$\alpha_2=1.3$, $\alpha_3=1$, $\alpha_4=-0.58$, 
$m^2_1=-25$ and $m^2_2=-6.1$.

The spectrum in the scalar sector is found diagonalizing the
$4\times 4$ matrix
\begin{equation}
M^2 = \left( \begin{array}{cccc}
2.8 & -1.0 & -5.8 & 4.3 \\
-1.0 & 0.77 & 2.6 & -2.0 \\
-5.8 & 2.6 & 91. & 6.7 \\
4.3 & -2.0 & 6.7 & 20. \\
\end{array}
\right)\;.
\label{mm}
\end{equation}
The eigenvalues give $m_1=9.6$, $m_2=4.6$,
$m_3=1.1$ and $m_4=0.56$. 
This particular case 
proves that there is SCPV in SUSY models containing two 
singlet fields.

\section{Conclusions} 

It is well known that the standard model and its 
minimal SUSY extension do not allow for SCPV.
To know whether in a more complicated Higgs sector
$CP$ can be broken spontaneously requires solving
equations that, in general, are difficult to handle.
We have shown that one can always build a combination
of triangles which is a solution of the minimum 
equations for the complex phases. 
In simpler cases (3D, 2D1S, $\rb$ and 4D models) 
the triangles are enough to solve also analytically 
the minimum equations for the moduli, and
in more complicated cases they help to find numerical solutions.

Using this method we have analyzed the possibility of SCPV
in SUSY models. 
Despite the large number of arbitrary parameters present
in these models, we find that a Higgs sector with only doublets 
does not provide SCPV:
if all the parameters in the Lagrangian are real, then
the Higgs VEVs can not be complex. 
This result has been proven
for the 4D model, the $\rb$ model 
with sneutrino VEVs (analogous to a 5D model), 
and the 6D model. 
In consequence, SUSY scenarios for SCPV require singlets. We find
that, if the singlets do not introduce dimensional parameters 
(i.e., no linear or bilinear terms in the superpotential), one
singlet is not enough to generate SCPV: the 2D1S and 
the 4D1S models have always real minima. The 
MSSM extended with two gauge singlets would be the minimal 
SUSY model where $CP$ violation can be generated spontaneously.

Other possibilities consistent with a soft origin of $CP$ violation
would be constrained by this result, like the 
4D model in Ref.~\cite{masi95a}. There  
all the parameters in the Lagrangian are taken 
real except for 
the mass terms of the Higgs fields. It is argued that 
these masses could appear at higher 
energy scales from large VEVs of
singlet fields  weakly coupled to the Higgs doublets,
and thus they can in principle be complex. Since
our analysis is  valid also in scenarios with a hierarchy between
singlet and doublet VEVs, it follows that the 
model at the large scale must
include at least two singlets. 
Other models that would
require at least two singlets to obtain soft $CP$ violation
within specific supersymmetric models can be found in 
\cite{moha97,fram97,eyal98}.

We would like to emphasize that our method to find
complex minima is not restricted to SUSY models. Any potential with all the
parameters real will give minimum equations for the 
phases of the VEVs that involve only sines (no cosines)
of the phases. In this case, 
using the sine law one can define a combination of
triangles that solve the equations. 
This is true for all supersymmetric potentials regardless of the
$CP$ properties of the fields, as well as for all nonsupersymmetric
potentials with $CP$ even fields and even for the physically interesting
minimal extension of the SM with one $CP$ odd real singlet.
Of course, there is always the trivial solution 
with all phases equal to zero and no SCPV. But the 
search for  complex minima beyond the simplest cases seems
almost unworkable  unless a method like the one described
in Section 2 is used.

\section*{Acknowledgments} 

We thank Alex Pomarol, Goran Senjanovi\'{c} and Atsushi Yamada
for helpful comments and discussions. M.M. thanks ICTP for its 
hospitality during the course of this work.
The work of M.M. was supported by CICYT under contract 
AEN96-1672 and by the Junta de Andaluc\'\i a under contract
FQM-101. The work of A.R. was supported in part by EEC grant
under the TMR contract ERBFMRX-CT960090. Part of this work
has been done during the ICTP Extended Workshop on Astroparticle 
Physics.

\newpage
\setlength{\unitlength}{1cm}
\begin{figure}[htb]
\begin{picture}(10,12)
\epsfxsize=10cm
\put(2,2.){\epsfbox{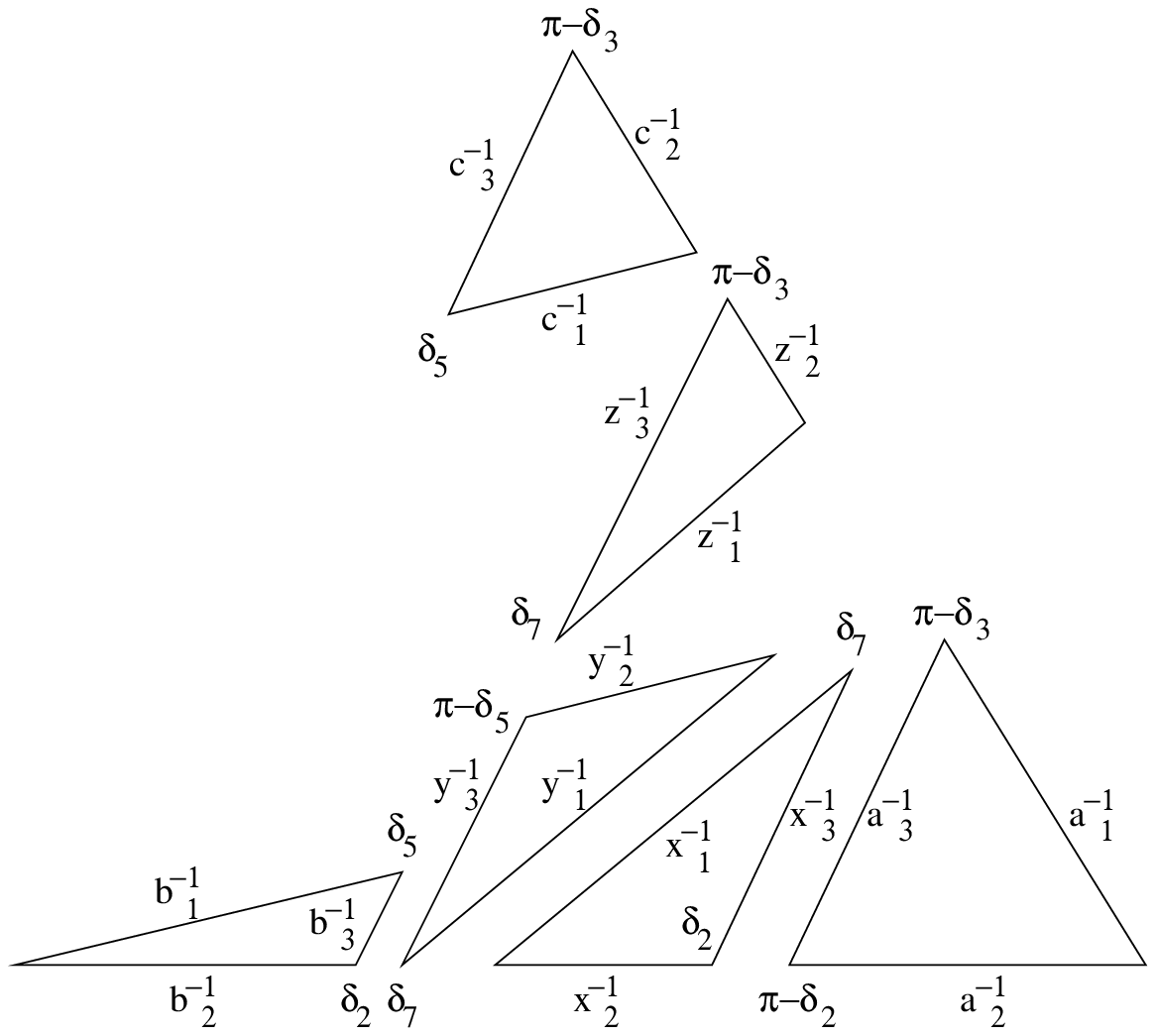}}
\end{picture}
\caption{The six triangles that make up the geometrical solution for the
minimum phase equations for the $\rb$ model. The lower right triangle is
the solution for the 3 Higgs doublet extension of the SM. The relations
between the sides of the triangles and the mass parameters are given in
the text.
\label{fig1}}
\end{figure}

\newpage
\begin{figure}[htb]
\begin{picture}(10,12)
\epsfxsize=12cm
\put(2,2.){\epsfbox{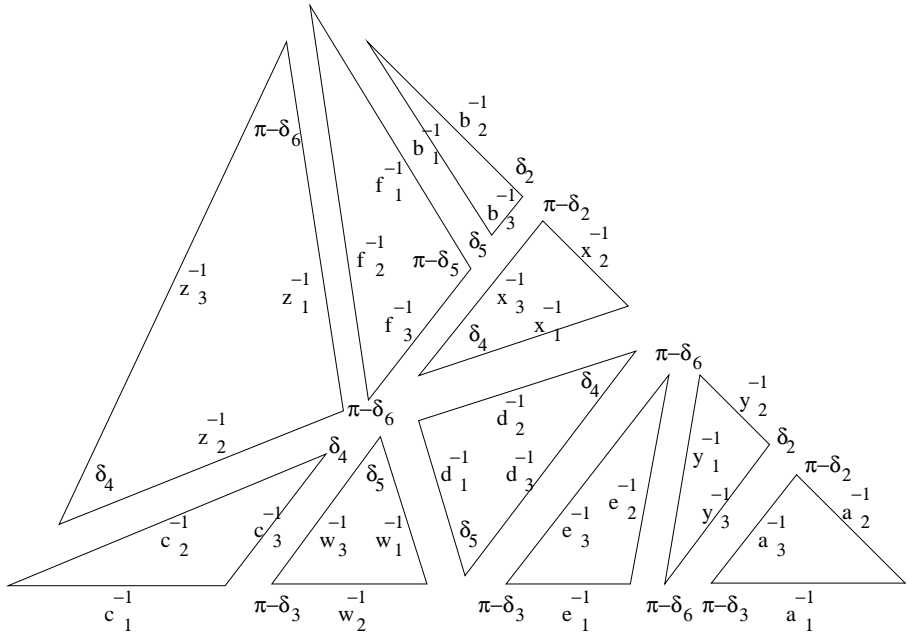}}
\end{picture}
\caption{6D model.
\label{fig2}}
\end{figure}

\newpage
\begin{figure}[htb]
\begin{picture}(10,12)
\epsfxsize=10cm
\put(2,2.){\epsfbox{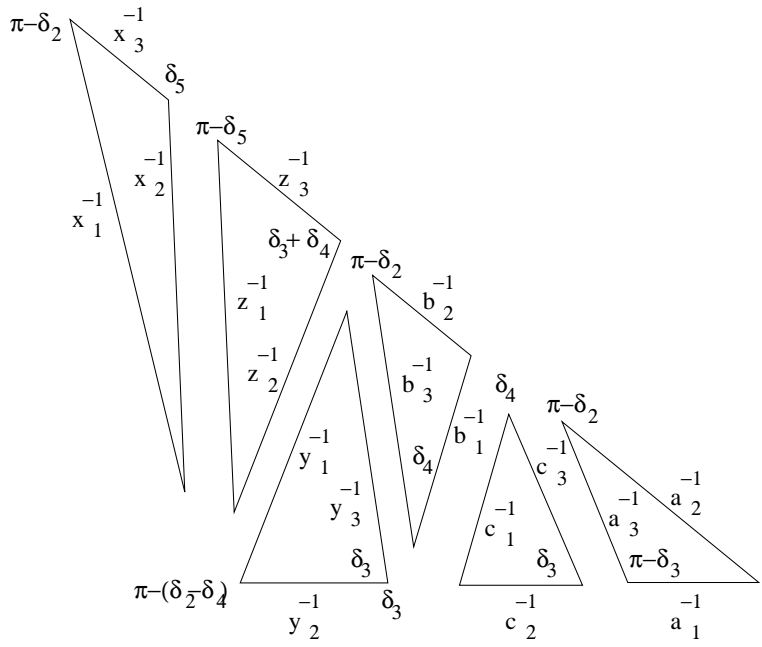}}
\end{picture}
\caption{4D1S model.
\label{fig3}}
\end{figure}

\newpage
\begin{figure}[htb]
\begin{picture}(10,12)
\epsfxsize=10cm
\put(2,2.){\epsfbox{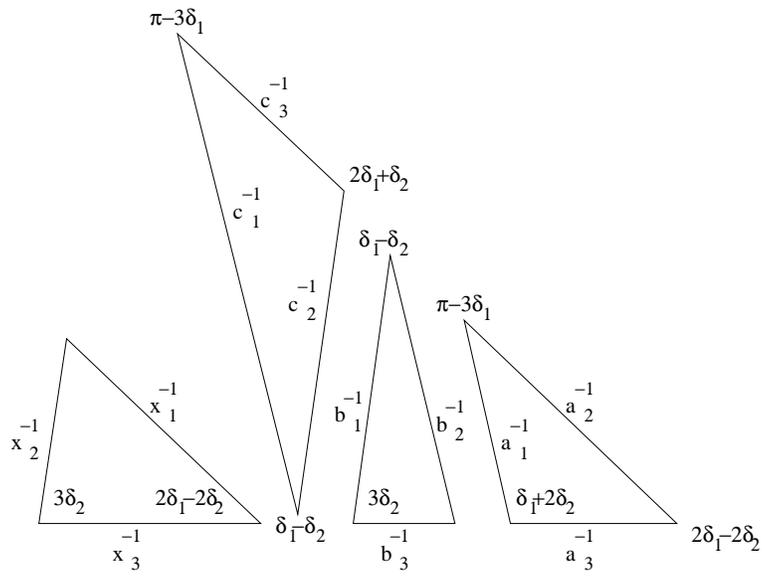}}
\end{picture}
\caption{2S model.
\label{fig4}}
\end{figure}

\end{document}